\documentclass[aps,pra,reprint,superscriptaddress,twocolumn]{revtex4-2}

\usepackage{xcolor}
\usepackage{amsmath}
\usepackage{graphicx}

\def\ket#1{\vert #1 \rangle}
\def\bra#1{\langle #1 \vert}

\def\kb#1{\vert #1 \rangle \!\langle #1 \vert}

\newcommand{\etal}{\textit{et al}.}

\DeclareMathOperator{\Tr}{Tr}

\begin{document}

\title{Cheating in quantum Rabin oblivious transfer using delayed measurements}

\author{James T. Peat}

\affiliation{SUPA, Institute of Photonics and Quantum Sciences, School of Engineering and Physical Sciences, Heriot-Watt University, Edinburgh EH14 4AS, United Kingdom}

\author{Erika Andersson}
\email[]{E.Andersson@hw.ac.uk}
\affiliation{SUPA, Institute of Photonics and Quantum Sciences, School of Engineering and Physical Sciences, Heriot-Watt University, Edinburgh EH14 4AS, United Kingdom}

\date{\today}

\begin{abstract}
Oblivious transfer has been the interest of study as it can be used as a building block for multiparty computation. There are many forms of oblivious transfer;  we explore a variant known as Rabin oblivious transfer. Here the sender Alice has one bit, and the receiver Bob obtains this bit with a certain probability. The sender does not know whether the receiver obtained the bit or not. For a previously suggested protocol, we show a possible attack using a delayed measurement. This allows a cheating party to pass tests carried out by the other party, while gaining more information than if they would have been honest. We show how this attack allows perfect cheating, unless the protocol is modified, and suggest changes which lower the cheating probability for the examined cheating strategies. 
\end{abstract}

\maketitle

\section{Introduction}

Quantum cryptography is by now well established, with quantum key distribution \cite{BB84} as the main application. Modern cryptography, however, encompasses more than message encryption and secret shared keys. There are numerous other cryptographic primitives, and quantum protocols for these, including for coin flipping \cite{doi:10.1137/14096387X,kitaev2003quantum}, bit commitment \cite{6108196} and identity authentication \cite{dutta2021shortreviewquantumidentity}. Here we examine a quantum protocol for a functionality known as oblivious transfer (OT). Oblivious transfer comes in different variants, including 1-out-of-2~\cite{10.1145/3812.3818,10.1145/1008908.1008920}, XOR~\cite{10.1007/3-540-69053-0_23} and Rabin oblivious transfer~\cite{Rabin2005HowTE}. These variants are related to each other; 1-out-of-2 and Rabin OT have been shown to be equivalent classically \cite{10.1007/3-540-48184-2_30}. It is currently unknown if the equivalency holds also for quantum protocols. Oblivious transfer has been of particular interest as it enables universal multiparty computation \cite{10.1145/62212.62215,10.1007/3-540-44750-4_9}. It has been shown \cite{PhysRevLett.78.3414,PhysRevA.56.1154}, however, that perfect information-theoretically secure one-sided two-party computation is impossible also quantum-mechanically, which implies that perfect quantum oblivious transfer is also impossible. Even so, quantum oblivious transfer is still studied, focusing either on realistic scenarios such as where cheating parties have access to bounded storage of quantum states~\cite{doi:10.1137/060651343}, or on finding bounds on cheating probabilities for sender and receiver~\cite{PRXQuantum.2.010335,10.5555/2481591.2481600}. He \etal\space\cite{PhysRevA.73.012331} propose a quantum protocol for Rabin oblivious transfer which they suggest is perfect and information-theoretically secure. They argue that since their protocol is not built from bit commitment, the proofs that disallow perfect OT no longer apply. Here we show that cheating is unfortunately possible in this protocol, using a delayed quantum measurement.

In Section \ref{Sec:Overview} we introduce Rabin oblivious transfer and review the protocol presented by He \etal. Then, in Section \ref{Sec:cheating} we outline the delayed measurement strategy. We then examine how this type of measurement can be used by both sender and receiver. First we explore when a malicious receiver Bob uses the strategy, then investigate a cheating sender Alice. This allows them both to cheat perfectly unless the protocol is modified. We then suggest how to modify the He protocol in order to reduce the cheating probability at least for the cheating strategies we examine. Cheating probabilities are still higher than in an ideal protocol. We do not prove that the cheating strategies we examine are the optimal ones, that is, cheating may still be possible with even higher probability.

\section{Overview of the Protocol}
\label{Sec:Overview}

\subsection{Rabin Oblivious Transfer}
Rabin oblivious transfer is a two-party cryptographic primitive involving a sender (Alice) and a receiver (Bob). Alice holds one bit which is sent to Bob, who receives this bit with a predetermined probability, often 1/2. In an ideal Rabin OT protocol Alice gains no information on whether Bob acquired the bit or not, and Bob gains no information about the bit value when he does not receive it.
Usually one defines Alice's cheating probability as the probability that she correctly guesses whether Bob obtained a bit or not. Bob's cheating probability is defined as the probability that he correctly guesses the value of Alice's bit.

\subsection{The protocol}
The protocol proposed by He \etal \cite{PhysRevA.73.012331} is rather complex with multiple steps. It requires the two parties to share many entangled states, on which they make measurements, discarding some of the states at various points throughout the protocol. It is the receiver Bob who creates the entangled states at the start of the protocol, sending parts of the entangled states to Alice. These states are each a four-qubit entangled state, with Alice receiving the first two qubits of each state. Alice then makes a choice between two different measurements for each state, then carries out these measurements on her part of the states she holds. These measurements have ``success” and ``failure” outcomes. The ``failure” outcomes are announced by Alice, and the corresponding states are discarded by both parties. There then follows verification by Bob, who selects a subset of the states which where not discarded, asking which measurement Alice made and its result. He then measures the part he holds to make sure that this is consistent with Alice's declaration, and also checks that Alice is selecting the two measurements with the correct probabilities. For the states that are not tested, Bob now chooses between two measurements with set probabilities for each state, and carries out the measurement on the part of the states he holds. Again, these measurements have ``success” and ``failure” outcomes, and Bob announces which measurements failed, and the corresponding states are discarded. Alice, just like Bob before, then checks a subset of the states to make sure that Bob has been honest, asking Bob to declare what measurement he made for each tested state, and its outcome. She then checks that this is consistent with the measurement result she obtained earlier. We call each of these a ``measure-verify block", with Alice measuring first, and Bob measuring second.

If neither party aborts in the verification steps, then Bob tries to project some of his states onto the Bell state $\ket{\Phi^+}=(\ket{00}+\ket{11})/\sqrt{2}$. He declares whether this is successful, and discards any states that fail to project onto $\ket{\Phi^+}$. Though we do not go into details, this prevents Alice from gaining an advantage in learning what measurement Bob made.
Finally Alice randomly selects one of the remaining states to carry out OT on. The choice of her measurement is then represented by the binary value $c$, with $0$ for one measurement choice and $1$ for the other. She then XORs $c$ with her bit $b$, sending the result $b'=b\oplus c$ to Bob. Depending on what measurement Bob chose to make, he will learn which measurement Alice made 50\% of the time, and therefore will obtain the bit half the time. This forms the outline of the Rabin protocol proposed by He \etal . 

It is the measure-verify block that we focus on, as a cheating party is successful if they know which measurement the other party chose to make.

\section{Cheating}
\label{Sec:cheating}
\subsection{The Generic Cheating Strategy}
The cheating strategy which a malicious party uses during the measure-verify block is similar for both Alice and Bob. It relies on a ``delayed" measurement strategy. An honest measuring party has to make a random decision between two measurements, with some probabilities. These measurements both have ``failure" outcomes. If a ``failure" outcome is obtained, then the measuring party announces this, and both parties discard the corresponding state. If the measurement is successful, the measuring party may later be asked to declare what result they obtained.
The security of the protocol rests on the fact that the measuring party has to make a choice between the two measurements, and has to actually make one of the measurements, in order to retain only the fraction of states for which either measurement gives a ``successful" outcome. If they try to delay making the measurement until after declaring for which states a ``successful" outcome was obtained, it would seem that they can no longer guarantee to obtain only ``successful" outcomes for the states that might be tested, and that therefore they would not be guaranteed to pass the tests by the other party.

\begin{figure}[t]
   \centering
        \includegraphics[width=\linewidth]{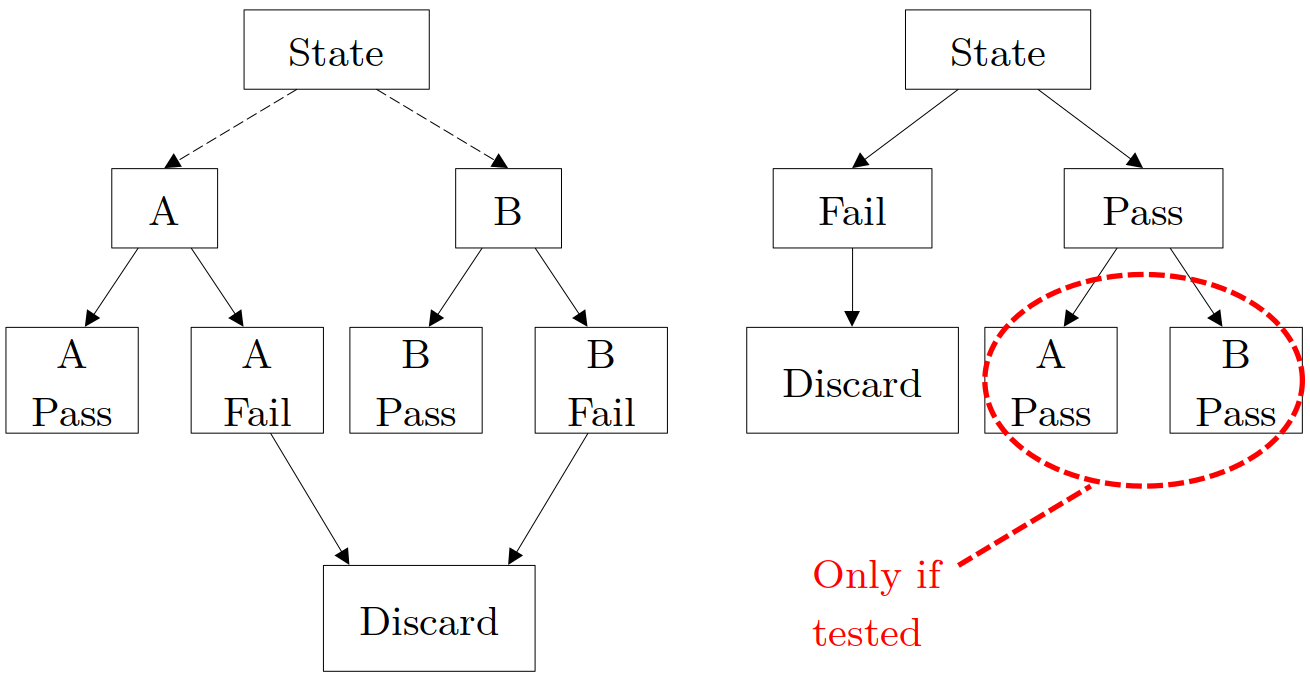}
    \caption{Block diagram of the possible measurements in the Rabin oblivious transfer protocol proposed by He \etal . The procedure used by an honest party is depicted on the left, and the procedure used by a dishonest party on the right. Here $A$ and $B$ represent the two different measurements that a measuring party can choose between.}
    \label{TreeDia}
\end{figure}

It is however possible to delay the choice of measurement, while still filtering out all states for which an ``unsuccessful" result is obtained at the earlier step where normally the choice between the two measurements should have been made.  We describe such a measurement that a cheating party can make, enabling them to gain more information than they otherwise should have. We build one combined measurement that no longer includes a choice. It is a two-outcome POVM with a ``failure" outcome and a ``success" outcome. These are a weighted sum of the measurement operators (for the measurements an honest party makes) corresponding to ``failure", and all the measurement operators corresponding to ``success", respectively. The sum is weighted according to the probabilities with which the measuring party would have chosen the two original measurements. If the measuring party obtains the failure outcome they, as before, declare and discard the state. If the measurement is ``success" then they continue the protocol. The verifying party will randomly select a fraction of the states that are not discarded. They will ask the measuring party what measurement they used and what the outcome was. When asked this, a cheating party can make a second measurement, which will return an outcome that is completely consistent with what they would have done if they had been honest, and they will be able to pass all tests. The construction of the measurement follows the results in ~\cite{PhysRevA.77.052104}, where it is shown that any generalised quantum measurement can be realized sequentially. The states that are carried forward have only been partially measured. The second measurement step that completes the measurement is only carried out if the state is tested. This allows a cheating party at the end of the protocol to make a measurement that gives a cheating probability which is higher than if they would have had to guess randomly. Fig. \ref{TreeDia} shows a diagram of the measurements made by an honest and a cheating party in the measure-verify blocks.
We now proceed to describe the measurements the cheating parties should make in more detail.

\subsection{Cheating Bob}
\label{Sub:BobCheating}
In a Rabin oblivious transfer protocol we define the cheating probability for the receiver Bob as the probability that he can correctly guess Alice's bit value. For a perfect protocol where Bob should obtain the bit value with probability $1/2$, Bob's cheating probability is $B_{ROT}=3/4$. This is because half the time Bob obtains the correct bit value, and when he does not, he then makes a random guess.

In the protocol by He \etal, a cheating Bob wants to learn which of the two measurements Alice made. Alice's choice is described by the binary value $c$. Bob wants to know $c$, as Alice encodes her bit $b$ in $b'=b\oplus c$ which is sent to Bob at the end of the protocol. This means that if Bob can work out $c$, then he is able to deduce $b$. When the protocol is implemented honestly, Bob's choice of measurement allows him to either find $c$, or gain no information on $c$. The verification stops Bob from always selecting the measurement that allows him to find $c$. An honest Bob measures by projecting into the basis
\begin{equation}
D_0 = \left\lbrace \ket{0}_+\ket{0}_+ , \ket{0}_+\ket{1}_+ ,  \ket{1}_+\ket{0}_+ ,     \ket{1}_+\ket{1}_+  \right\rbrace,
\label{d0}
\end{equation}
with probability $2/3$ and into the basis
\begin{equation}
D_1 = \left\lbrace \ket{0}_\times\ket{0}_\times , \ket{0}_\times\ket{1}_\times ,  \ket{1}_\times\ket{0}_\times ,     \ket{1}_\times\ket{1}_\times  \right\rbrace,
\label{d1}
\end{equation} 
with probability $1/3$, where $\ket{x}_\times = \left(\ket{0}_++(-1)^x\ket{1}_+\right)/\sqrt{2}$. The last two states of each of these measurement bases are regarded as failure outcomes, meaning that the state will be discarded. We can therefore use the first two states in each measurement basis to construct the measurement operator for our new POVM, corresponding to ``success". This measurement operator is given by
\begin{equation}
\Pi_s= \frac{2}{3}(\Pi_{00+}+\Pi_{11+})+\frac{1}{3}(\Pi_{00\times}+\Pi_{11\times}),
\end{equation}
where
\begin{equation}
\Pi_{ab+}=\ket{a}_+\ket{b}_+ \bra{a}_+\bra{b}_+,
\end{equation}
and correspondingly for $\Pi_{ab\times}$, where $a,b = 0,1$.
We do not need to look in detail at the ``failure" measurement operator, since if this outcome is obtained, the state is discarded. It would however be constructed analogously. When Bob obtains the ``success" outcome and Alice selects the state to test, Bob makes a second measurement, constructed using the method in~\cite{PhysRevA.77.052104}, which is entirely consistent with what he would have done if honest. 
This second measurement would also give Bob knowledge of what measurement he would have chosen and its outcome if he had been honest, allowing him to pass Alice's verification. Bob's state after the measurement is found by applying the Kraus operator $m_i$ corresponding to the measurement outcome he obtained. The Kraus operators satisfy
\begin{equation}
\Pi_i=m_i^\dagger m_i.
\end{equation}
The state, conditioned on the outcome corresponding to $\Pi_i$, is transformed as
\begin{equation}
\ket{\varphi}\rightarrow\ket{\varphi^\prime}=\frac{m_i\ket{\varphi}}{\sqrt{p_i}},
\label{PostMesur}
\end{equation}
where $p_i$ is the probability of the outcome $i$. The Kraus operators are not unique. However, different choices only differ in a unitary operation which does not matter here (it can be seen as a unitary operation, conditioned on the outcome, applied to the post-measurement state, and does not affect the information Bob obtains). We therefore select 
\begin{equation}
m_s=\sqrt{\Pi_s}
\label{kraus}
\end{equation}
as the Kraus operator for the ``success" outcome. To find this Kraus operator we first write the POVM element in matrix form using the computational basis,
\begin{equation}
\Pi_s=\frac{1}{6}
\left(\begin{array}{cccc}
5 & 0 & 0 & 1	\\
0 & 1 & 1 & 0	\\
0 & 1 & 1 & 0	\\
1 & 0 & 0 & 5
\end{array}\right).
\end{equation}
We can then diagonalise this to find its eigenvalues
\begin{equation}
\lambda_1=1, \quad \lambda_2=2/3, \quad \lambda_3=1/3, \quad \lambda_4=0.
\end{equation}
The eigenvectors are
\begin{equation}
\begin{array}{c}
\ket{\lambda_1}=\frac{1}{\sqrt{2}}(\ket{00}+\ket{11})=\ket{\Phi^+},
\\
\ket{\lambda_2}=\frac{1}{\sqrt{2}}(\ket{00}-\ket{11})=\ket{\Phi^-}, 	\\
\ket{\lambda_3}=\frac{1}{\sqrt{2}}(\ket{01}+\ket{10})=\ket{\Psi^+},
\\
\ket{\lambda_4}=\frac{1}{\sqrt{2}}(\ket{01}-\ket{10})=\ket{\Psi^-}. 
\end{array}
\end{equation}
This allows us to rewrite $\Pi_s$ as $\Pi_s=\sum_i\lambda_i\kb{\lambda_i}$, which means that we can express the Kraus operator as
\begin{equation}
m_s = \kb{\Phi^+}+\sqrt{\frac{2}{3}}\kb{\Phi^-}+\sqrt{\frac{1}{3}}\kb{\Psi^+}.
\end{equation}
Since Bob is the second to carry out the measure-verify procedure and we assume that Alice is honest when she carries out the first measurement, there are two possible states corresponding to $c=0$ and $c=1$.
We describe the total state held by Alice and Bob as $\ket{\psi^{(c)}}$, with the subscripts denoting the parts Alice and Bob hold. These states are written as
\begin{equation}
\begin{split}
\ket{\psi^{(0)}}=&\frac{1}{2\sqrt{2}}((\ket{\Phi^-}-\ket{\Psi^+})_A(\ket{\Phi^+}-
\ket{\Psi^+})_B
\\
+&(\ket{\Phi^-}+\ket{\Psi^+})_A(\ket{\Phi^-}-\ket{\Psi^-})_B),
\end{split}
\end{equation}
\begin{equation}
\begin{split}
\ket{\psi^{(1)}}&=\frac{1}{2\sqrt{2}}(   (\ket{0}_\times\ket{0}_+ \!-\! \ket{1}_\times \ket{1}_+)_A(\ket{\Phi^+}+\ket{\Psi^+})_B \\
 &+   (\ket{0}_\times \ket{0}_+ + \ket{1}_\times \ket{1}_+)_A(\ket{\Phi^-}-\ket{\Psi^-})_B).
\end{split}
\end{equation}
This is a bipartite state shared between Alice and Bob. When Bob makes his measurement, the whole state transforms with $\hat{I}_A\otimes m_{s,B}$, where $\hat I_A$ is the identity operator on Alice's system. Using this and Eq. (\ref{PostMesur}), we find the post-measurement states
\begin{equation}
\begin{split}
\ket{\psi^{\prime\,0}}&=\frac{\hat{I}_A\otimes m_{s,B}\ket{\psi^{(0)}}}{\sqrt{p_s}} \\
&=\frac{1}{2}[(\ket{\Phi^-}-\ket{\Psi^+})_A(\ket{\Phi^+}-\sqrt{\frac{1}{3}}\ket{\Psi^+})_B\\
&+\sqrt{\frac{2}{3}}(\ket{\Phi^-}+\ket{\Psi^+})_A\ket{\Phi^-}_B],
\end{split}
\end{equation}
and
\begin{equation}
\begin{split}
&\ket{\psi^{\prime\,1}}=\frac{\hat{I}_A\otimes m_{s,B}\ket{\psi^{(1)}}}{\sqrt{p_s}} \\
&=\frac{1}{2}[(\ket{0}_\times\ket{0}_+\!-\ket{1}_\times\ket{1}_+)_A(\ket{\Phi^+}+\sqrt{\frac{1}{3}}\ket{\Psi^+})_B \\
&+\sqrt{\frac{2}{3}}(\ket{0}_\times\ket{0}_++\ket{1}_\times\ket{1}_+)_A\ket{\Phi^-}_B].
\end{split}
\end{equation}
Here we look at what Bob holds for each state. For this we need the reduced density matrix. The density matrix of the total state is described by
\begin{equation}
\rho_{AB}^c=\kb{\psi^{\prime c}}.
\end{equation} 
Bob's system can then be described by tracing out Alice,
\begin{equation}
\rho_B^c=\Tr_A(\rho_{AB}^c).
\end{equation}
The two possible states that Bob could hold are
\begin{equation}
\label{eq:Bobstwostates}
\begin{split}
\rho_B^0&=\frac{1}{4}(\kb{x_0}+\kb{y_0}),\\
\rho_B^1&=\frac{1}{4}(\kb{x_1}+\kb{y_1}),
\end{split}
\end{equation}
with
\begin{equation}
\begin{split}
\ket{x_0}&=\ket{\Phi^+}-\sqrt{\frac{1}{3}}\ket{\Psi^+}+\sqrt{\frac{2}{3}}\ket{\Phi^-},\\
\ket{y_0}&=\sqrt{\frac{2}{3}}\ket{\Phi^-}+\sqrt{\frac{1}{3}}\ket{\Psi^+}-\ket{\Phi^+},\\
\ket{x_1}&=\ket{\Phi^+}+\sqrt{\frac{1}{3}}\ket{\Psi^+}+\sqrt{\frac{2}{3}}\ket{\Phi^-},\\
\ket{y_1}&=\ket{\Phi^+}+\sqrt{\frac{1}{3}}\ket{\Psi^+}-\sqrt{\frac{2}{3}}\ket{\Phi^-}.
\end{split}
\end{equation}
At this point in the protocol, an honest Bob makes a projection onto $\ket{\Phi^+}$, and discards the state if he fails. If he is cheating, he could use this step to his advantage, and make an unambiguous measurement to distinguish between the above states. If he is able to tell Alice to discard the state whenever this measurement fails, then he could use this to cheat perfectly. If Alice monitors how often Bob asks her to discard the state -- which is not part of the protocol as described by He \etal~ -- then this will limit Bob's cheating probability. We will first show that even if Bob discards no states at all, his optimal minimum-error measurement allows him to cheat. After that we will show how Bob could cheat perfectly using an unambiguous measurement, and finally give a bound for how often Bob can cheat if Alice monitors the probability that he discards the state.

We can find the probability for Bob to distinguish between the two states in Eq. \ref{eq:Bobstwostates}, which is what he needs to do in order to learn the value of $c$ and cheat successfully. Since there are only two states, we can use the standard result for success probability of the minimum-error measurement, the so-called Helstrom measurement \cite{alma9998903502466},
\begin{equation}
p_s=\frac{1}{2}(1+\Tr(|p_0\rho_0-p_1\rho_1|)),
\label{TwostateMEM}
\end{equation}
where $|A|=\sqrt{A^\dagger A}$. This gives Bob a cheating probability of
\begin{equation}
B_{OT}=\frac{1}{6}(3+\sqrt{3})\approx0.789.
\end{equation}
This is greater than $0.75$ which, as discussed before, would be what a perfect protocol achieves. 

Next we examine the step where the state is projected onto $\ket{\Phi^+}$, to further increase Bob's cheating probability. This step allows him, when cheating, to discard states after making a measurement. This means that Bob can make an unambiguous state discrimination measurement. To find the corresponding measurement operators we use the form in \cite{PhysRevLett.96.070401} for finding the maximum confidence measurement, since whenever unambiguous state discrimination is possible, the maximum confidence measurement is also an unambiguous state discrimination measurement. To simplify calculations we switch to vector notation with the basis
\begin{equation}
\ket{\Phi^+}=\left(
\begin{array}{c}
1\\
0\\
0
\end{array}
\right)\!\!,
\ 
\ket{\Psi^+}=\left(
\begin{array}{c}
0\\
1\\
0
\end{array}
\right)\!\!,
\ 
\ket{\Phi^-}=\left(
\begin{array}{c}
0\\
0\\
1
\end{array}
\right)\!\!.
\end{equation}
Using these basis states, the elements of the POVM are given by 
\begin{equation}
\Pi_0=\frac{1}{6}
\left(\!\!
\begin{array}{ccc}
1&-\sqrt{3}&0\\
-\sqrt{3}&3&0\\
0&0&0
\end{array}
\!\!\right)\!\!,
\ 
\Pi_1=\frac{1}{6}
\left(\!\!
\begin{array}{ccc}
1&\sqrt{3}&0\\
\sqrt{3}&3&0\\
0&0&0
\end{array}
\!\!\right)\!\!,
\end{equation}
\begin{equation}
\Pi_?=
\left(
\begin{array}{ccc}
\frac{2}{3}&0&0\\
0&0&0\\
0&0&1
\end{array}
\right)\!\!.
\end{equation}
The probability of each outcome is calculated using $p_c(\rho)=\Tr(\Pi_c\rho)$, which gives
\begin{equation}
\begin{array}{ccc}
p_0(\rho_B^0)=\frac{1}{3},&\quad p_1(\rho_B^0)=0,&\quad p_?(\rho_B^0)=\frac{2}{3},
\\ \\
p_0(\rho_B^1)=0,&\quad p_1(\rho_B^1)=\frac{1}{3},&\quad p_?(\rho_B^1)=\frac{2}{3}.
\end{array}
\end{equation}
With this Bob can now cheat perfectly ($B_{OT}=1$), since whenever he obtains an inconclusive outcome, he tells Alice to discard the state. This is because as stated, the protocol does not limit the fraction Bob can discard. 

We can add the requirement that the fraction Bob can discard is $1/3$. This is the probability that in the honest case he would discard a state. To obtain a rough bound on Bob's cheating probability in this case, we can use the minimum-error measurement half the time, and unambiguous state discrimination the remaining time. This gives a probability of $1/3$ to discard the state. Bob's cheating probability is then
\begin{equation}
B_{OT}=\frac{1}{2}\times1+\frac{1}{2}\times\frac{1}{6}(3+\sqrt{3})\approx0.894.
\end{equation}
This is not Bob's optimal measurement for cheating. The optimal measurement would be a measurement where Bob obtains the inconclusive result with a fixed probability, while minimising the error of incorrectly identifying the state \cite{PhysRevA.86.032314,PhysRevA.86.040303}.
Bob could also try cheat at the very beginning by sending a different state to Alice. In the cheating strategy discussed here he is honest until he makes his measurements. However, this more general cheating strategy could only increase Bobs cheating probability. We have thus shown that Bob cheats with at least probability $B_{OT}\approx 0.894$ in the He protocol.

\subsection{Alice Cheating}
\label{Sub:AliceCheating}
In a Rabin oblivious transfer protocol we define Alice's cheating probability as the probability that she correctly guesses whether Bob received a bit or not.
For an ideal protocol where Bob receives the bit with probability $1/2$, this probability is $A_{ROT}=0.5$. This is because all Alice can do is make a random guess where she would be correct half the time.

In the protocol we are examining, a cheating Alice wants to learn which of the two measurements Bob made. Bob's choice is described by the binary value $d$. If $d=1$ then Bob receives the bit, otherwise he does not. When an honest Alice measures, she projects the two qubits she holds either into the Bell basis
\begin{equation}
C_0 = \{\ket{ \Phi^+},\ket{\Phi^-},\ket{\Psi^+},\ket{\Psi^-}\},
\end{equation}
or the basis
\begin{equation}
C_1 = \{\ket{0}_\times\ket{0}_+,\ket{0}_\times\ket{1}_+,\ket{1}_\times\ket{0}_+,\ket{1}_\times\ket{1}_+\},
\end{equation}
with probability 1/2 each. Alice's projection is successful when she obtains either $\ket{\Phi^-}$, $\ket{\Psi^+}$, $\ket{0}_\times\ket{0}_ +$ or $\ket{0}_\times\ket{1}_+$. These elements are used to build Alice's measurement operator corresponding to ``success", which is given by
\begin{equation}
\Pi_s=\frac{1}{2}(\Pi_{\ket{\Phi^-}}+\Pi_{\ket{\Psi^+}}+\Pi_{\ket{0}_\times\ket{0}_+}+\Pi_{\ket{1}_\times\ket{1}_+}),
\end{equation}
where $\Pi_{\ket{x}}=\kb{x}$. To calculate the resulting state after this measurement, we again find the corresponding Kraus operator using Eq. (\ref{kraus}). In matrix notation, using the computational basis, the ``success" measurement operator is given by
\begin{equation}
\Pi_s=
\frac{1}{4}
\left(\begin{array}{cccc}
2 & 0 & 1 & -1	\\
0 & 2 & 1 & -1	\\
1 & 1 & 2 & 0	\\
-1 & -1 & 0 & 2
\end{array}\right).
\end{equation}
This has the eigenvalues
\begin{equation}
\lambda_1=1, \quad \lambda_2=1/2, \quad \lambda_3=1/2, \quad \lambda_4=0,
\end{equation}
and the corresponding eigenvectors
\begin{equation}
\begin{gathered}
\ket{\lambda_1}=\frac{1}{2}(\ket{00}+\ket{01}+\ket{10}-\ket{11}), 
\\
\ket{\lambda_2}=\frac{1}{\sqrt{2}}(\ket{10}+\ket{11}),
\quad
\ket{\lambda_3}=\frac{1}{\sqrt{2}}(\ket{00}-\ket{01}), 
\\
\ket{\lambda_4}=\frac{1}{2}(\ket{00}+\ket{01}-\ket{10}+\ket{11}). 
\end{gathered}
\end{equation}
This allows us to express the Kraus operator as
\begin{equation}
m_s=\kb{\lambda_1}+\sqrt{\frac{1}{2}}(\kb{\lambda_2}+\kb{\lambda_3}).
\end{equation}
Since Alice is the first party to make a measurement, the state she measures, if it has been honestly prepared by Bob according to the protocol, is
\begin{equation}
\begin{split}
\ket{\psi} = (&\ket{00}_{+A}\ket{00}_{+B}+\ket{11}_{+A}\ket{01}_{+B}
\\
+&\ket{00}_{\times A}\ket{10}_{+B}+\ket{11}_{\times A}\ket{11}_{+B})/2,
\end{split}
\end{equation}
with $\ket{00}_{sX}=\ket{0}_{sX}\ket{0}_{sX}$.
The state after Alice has obtained a ``success" outcome is then given by
\begin{equation}
\ket{\psi^\prime}=\sqrt{2}m_{s,A}\otimes\hat{I}_B\ket{\psi}.
\end{equation}
This is the total state held by Alice and Bob, which an honest Bob will then go on to make measurements on. 
Bob's subsequent measurement is a projection on the part he holds into a basis $D_0$ or $D_1$. Again, we only care about the outcomes that are not discarded, because the discarded states do not play a further part in the protocol. The four ``success" outcomes of Bob's measurement correspond to the states $\ket{0}_+\ket{0}_+ $, $\ket{1}_+\ket{1}_+$, $\ket{0}_\times\ket{0}_\times $, $ \ket{1}_\times\ket{1}_\times$. We denote the resulting states after the measurement by $\ket{\psi^\prime_{00}},\ket{\psi^\prime_{11}},\ket{\psi^\prime_{0\times}},\ket{\psi^\prime_{1\times}}$ respectively, given by
\begin{equation}
\begin{split}
&\ket{\psi^\prime_{00}}=2\hat{I}_A\otimes\Pi_{\ket{0}_+\ket{0}_+,B}\ket{\psi^\prime}
=\\
&\frac{1}{2\sqrt 2}\left[({1+\sqrt 2})\ket{00}+(1-\sqrt 2)\ket{01}+\ket{10}-\ket{11}\right]_A\otimes\ket{00}_B,
\\
&\ket{\psi^\prime_{11}}=2\hat{I}_A\otimes\Pi_{\ket{1}_+\ket{1}_+,B}\ket{\psi^\prime}
=\\
&\frac{1}{2\sqrt{2}}\left[(1-\sqrt{2})\ket{00}+(1+\sqrt{2})\ket{01}+\ket{10}-\ket{11}\right]_A\otimes\ket{11}_B,\\
&\ket{\psi^\prime_{0\times}}=2\hat{I}_A\otimes\Pi_{\ket{0}_\times\ket{0}_\times,B}\ket{\psi^\prime}
=\\
&\frac{1}{4}(\ket{00}-\ket{01}+\ket{10}+\ket{11})_A\otimes(\ket{00}+\ket{01}+\ket{10}+\ket{11})_B,\\
&\ket{\psi^\prime_{1\times}}=2\hat{I}_A\otimes\Pi_{\ket{1}_\times\ket{1}_\times,B}\ket{\psi^\prime}
=\\
&\frac{1}{4}(\ket{00}-\ket{01}-\ket{10}-\ket{11})_A\otimes(\ket{00}-\ket{01}-\ket{10}+\ket{11})_B.
\end{split}
\end{equation}
For the case where Bob has $d=0$, he performs a further projection onto $\ket{\Phi^+}$ discarding any states that fail to project onto this state. This further projection gives 
\begin{equation}
\begin{split}
\ket{\psi^{\prime\prime}_{00}}=&\sqrt{2}\hat{I}_A\otimes\Pi_{\ket{\Phi^+},B}\ket{\psi^\prime_{00}}\\
=&\frac{1}{2\sqrt{2}}[(1+\sqrt{2})\ket{00}+(1-\sqrt{2})\ket{01}\\ 
&~~~~~~~~~~~~~~~~+\ket{10}-\ket{11}]_A\otimes\ket{\Phi^+}_B\\
\ket{\psi^{\prime\prime}_{11}}=&\sqrt{2}\hat{I}_A\otimes\Pi_{\ket{\Phi^+},B}\ket{\psi^\prime_{11}}
\\
=&\frac{1}{2\sqrt{2}}[(1-\sqrt{2})\ket{00}+(1+\sqrt{2})\ket{01}\\ 
&~~~~~~~~~~~~~~~+\ket{10}-\ket{11}]_A\otimes\ket{\Phi^+}_B.
\end{split}
\end{equation}
From Alice's point of view, the state could be either
\begin{equation}
\begin{split}
\ket{\psi^{\prime\prime}_{00}}_A&=\frac{1}{2\sqrt{2}}((1+\sqrt{2})\ket{00}+(1-\sqrt{2})\ket{01}+\ket{10}-\ket{11}),\\
\ket{\psi^{\prime\prime}_{11}}_A&=\frac{1}{2\sqrt{2}}((1-\sqrt{2})\ket{00}+(1+\sqrt{2})\ket{01}+\ket{10}-\ket{11}),
\end{split}
\end{equation}
for $d=0$, and
\begin{equation}
\begin{split}
\ket{\psi^\prime_{0\times}}_A&=\frac{1}{2}(\ket{00}-\ket{01}+\ket{10}+\ket{11}),\\
\ket{\psi^\prime_{1\times}}_A&=\frac{1}{2}(\ket{00}-\ket{01}-\ket{10}-\ket{11}),
\end{split}
\end{equation}
for $d=1$.
Therefore the two states Alice has to distinguish between are 
\begin{equation}
\begin{split}
\rho_A^0&=\frac{1}{2}(\ket{\psi^{\prime\prime}_{00}}_A{}_A\bra{\psi^{\prime\prime}_{00}}+\ket{\psi^{\prime\prime}_{11}}_A{}_A\bra{\psi^{\prime\prime}_{11}}),\\
\rho_A^1&=\frac{1}{2}(\ket{\psi^{\prime}_{0\times}}_A{}_A\bra{\psi^{\prime}_{0\times}}+\ket{\psi^{\prime}_{1\times}}_A{}_A\bra{\psi^{\prime}_{1\times}}).
\end{split}
\end{equation} 
Alice's success probability for distinguishing between these two states is found using Eq. (\ref{TwostateMEM}), and is equal to $A_{ROT}=3/4$, which is greater than $1/2$, which is what it would be for an ideal protocol. This however is still not Alice's optimal measurement. When Alice is honest, she randomly selects one of the remaining states to carry out OT on. The fact that she makes a choice, and discards all but one of the states, allows a cheating Alice to instead perform unambiguous state discrimination. When her unambiguous measurement is successful, she knows for certain which state she held. When it fails, she knows not to select that state. 

The quantum system Alice holds is actually three-dimensional, but so far described in a four-dimensional space. To simplify finding the measurement operators for unambiguous state discrimination, we describe the system in three dimensions. This can be done by first writing $\rho_A^1$ in its eigenbasis,
\begin{equation}
\rho_A^1=\frac{1}{2}(\kb{\phi_0}+\kb{\phi_1}),
\end{equation}
with
\begin{equation}
\ket{\phi_0}=\frac{1}{\sqrt{2}}(\ket{00}-\ket{01}),~
\ket{\phi_1}=\frac{1}{\sqrt{2}}(\ket{10}+\ket{11}).
\end{equation}
Similarly, we have
\begin{equation}
\rho_A^0=\frac{1}{2}(\kb{\phi_0}+\kb{\phi_2}),
\end{equation}
with
\begin{equation}
\ket{\phi_2}=\frac{1}{2}(\ket{00}+\ket{01}+\ket{10}-\ket{11}).
\end{equation}
The states $\{\ket{\phi_0},\ket{\phi_1},\ket{\phi_2}\}$ form an orthonormal set and therefore a basis. The supports for the states $\rho_A^0$ and $\rho_A^1$ each contain one unique basis state, and share the third. It immediately follows that the unambiguous state discrimination measurement is to project into this basis. The outcome corresponding to $\ket{\phi_0}$ is the inconclusive result, and $\ket{\phi_2}$, $\ket{\phi_1}$ correspond to $\rho_A^0$ and $\rho_A^1$ respectively. Half the time, this measurement correctly identifies the state, and the rest of the time the result is inconclusive. Therefore, for half of the states, Alice can learn Bob's value for $d$ with certainty. She selects one of these states for the oblivious transfer, meaning that Alice cheats perfectly ($A_{ROT}=1$). Not only can she cheat perfectly, she can force which outcome Bob obtains by selecting the state with the corresponding result.

\subsection{Improving the protocol}
To help protect against this strategy we can make a few suggestions to reduce the cheating bound. The first stops Alice from cheating perfectly with the delayed measurement. If we remove the last step so Alice no longer selects one state for OT which produces a single instance of oblivious transfer. The protocol will now  produce many instances of oblivious transfer. This change means that Alice can no longer make an unambiguous state discrimination measurement, stopping her from cheating perfectly. When the protocol is modified in this way, Alice's best strategy in the final step becomes a minimum-error measurement, since each remaining state will be used for an instance of Rabin oblivious transfer. This minimum-error probability was calculated in Section \ref{Sub:AliceCheating}, giving Alice a cheating probability of $A_{ROT}=3/4$.

The other protocol modification we suggest stops Bob from being able to cheat perfectly, and was mentioned in Section \ref{Sub:BobCheating}. We suggested that Alice should monitor the fraction of states Bob discards when he projects onto $\ket{\Phi^+}$. Again this stops Bob from making an unambiguous state discrimination measurement, and he has to instead make a measurement with a fixed rate of failure. With this modification we found a lower bound on Bob's cheating probability of $B_{ROT}\approx0.894$. There may be better cheating strategies for Bob.

We highlight here that we have only examined specific cheating strategies, and hence the cheating probabilities we have found are only lower bounds. As mentioned in Section \ref{Sub:BobCheating}, Bob could make a better measurement. We have also only considered Bob cheating from his first measurement, acting honestly before then. Since Bob prepares the initial state he may be able to cheat further by sending a different state.

\section{Conclusions}
We examined a proposed quantum Rabin oblivious transfer protocol, exploring how both sender and receiver can cheat using delayed measurements. We showed how a cheating party can pass tests by the other party without being detected. We also showed how, when not tested, the cheating strategy allowed a cheating party to obtain more information on the state than when honest, leading to a cheating probabilty that is larger than for an ideal protocol.

If the protocol is carried out as suggested by He \etal, both sender and receiver could cheat perfectly using the cheating strategies we have examined. We made suggestions for how to modify the protocol so as to reduce the bounds on the cheating probabilities.
Nevertheless, we only examined specific cheating strategies, and it may be possible for in particular the receiver Bob to further increase his cheating probability by preparing a different initial entangled state.
All in all, neither the original protocol nor its modified variant are perfect. It remains an open question what the lowest possible cheating probabilities are in quantum Rabin oblivious transfer, and whether or not Rabin and 1-out-of-2 oblivious transfer are equivalent in the quantum setting.

\section{Acknowledgments}
This work was supported by the UK Engineering and Physical Sciences Research Council (EPSRC) under Grant No. EP/T001011/1.

\bibliography{CheatinginROT}

\end{document}